\documentclass[prl,nobalancelastpage,superscriptaddress,twocolumn,nofootinbib]{revtex4-1}

\usepackage{graphicx}
\usepackage{amsmath, amsxtra, amssymb, mathtools, nccmath, xspace}
\usepackage[utf8]{inputenc}
\usepackage{float}
\usepackage{placeins} 
\usepackage{xcolor} 
\usepackage{nicefrac} 
\usepackage{xspace}     

\begin{document}
\title{
Generation and manipulation of Schr\"odinger cat states in Rydberg atom arrays
}

\newcommand{\Harvard}{Department of Physics, Harvard University, Cambridge, MA
02138, USA}
\newcommand{\Gordon}{Department of Physics, Gordon College, Wenham, MA 01984, USA}
\newcommand{\Chicago}{Institute for Molecular Engineering, University of
Chicago, Chicago, IL 60637, USA}
\newcommand{\ITAMP}{ITAMP, Harvard-Smithsonian Center for Astrophysics,
Cambridge, MA 02138, USA}
\newcommand{\Berkeley}{Department of Physics, University of California
Berkeley, Berkeley, CA 94720, USA}
\newcommand{\Julich}{Forschungszentrum J\"ulich, Institute of Quantum Control (PGI-8), D-52425 J\"ulich, Germany}
\newcommand{\Ulm}{Institute for Quantum Optics and IQST Center, Universit\"at Ulm, D-89081 Ulm, Germany}
\newcommand{\Padova}{Dipartimento di Fisica e Astronomia "G. Galilei",
Universit\`a degli Studi di Padova \& INFN, I-35131 Italy}
\newcommand{\Koln}{Institute for Theoretical Physics, University of Cologne, D-50937 Cologne, Germany}
\newcommand{\Caltech}{Division of Physics, Mathematics and Astronomy,
California Institute of Technology, Pasadena, CA 91125, USA}
\newcommand{\MIT}{Department of Physics and Research Laboratory of Electronics,
Massachusetts Institute of Technology, Cambridge, MA 02139, USA}

\author{A. Omran}
\thanks{These authors contributed equally to this work}
\affiliation{\Harvard}
\author{H. Levine}
\thanks{These authors contributed equally to this work}
\affiliation{\Harvard}
\author{A. Keesling}
\author{G. Semeghini}
\affiliation{\Harvard}
\author{T. T. Wang}
\affiliation{\Harvard}
\affiliation{\Gordon}
\author{S. Ebadi}
\affiliation{\Harvard}
\author{H. Bernien}
\affiliation{\Chicago}
\author{A. S. Zibrov}
\affiliation{\Harvard}
\author{H. Pichler}
\affiliation{\Harvard}
\affiliation{\ITAMP}
\author{S. Choi}
\affiliation{\Berkeley}
\author{J. Cui}
\affiliation{\Julich}
\author{M. Rossignolo}
\affiliation{\Ulm}
\author{P. Rembold}
\affiliation{\Julich}
\author{S. Montangero}
\affiliation{\Padova}
\author{T. Calarco}
\affiliation{\Julich}
\affiliation{\Koln}
\author{M. Endres}
\affiliation{\Caltech}
\author{M. Greiner}
\affiliation{\Harvard}
\author{V. Vuleti\'c}
\affiliation{\MIT}
\author{M. D. Lukin}
\thanks{To whom correspondence should be addressed; E-mail: lukin@physics.harvard.edu}
\affiliation{\Harvard}

\begin{abstract}
    Quantum entanglement involving coherent superpositions of macroscopically
    distinct states is among the most striking features of quantum theory, but
    its realization is challenging because such states are extremely fragile.
    Using a programmable quantum simulator based on neutral atom arrays with
    interactions mediated by Rydberg states, we demonstrate the creation
    of ``Schr\"odinger cat" states of the Greenberger-Horne-Zeilinger (GHZ)
    type with up to 20 qubits. Our approach is based on engineering the
    energy spectrum and using optimal control of the many-body system. We
    further demonstrate entanglement manipulation by using GHZ states to
    distribute entanglement to distant sites in the array, establishing
    important ingredients for quantum information processing and quantum
    metrology.
\end{abstract}

\maketitle

Greenberger-Horne-Zeilinger (GHZ) states constitute an important class of
entangled many-body states~\cite{Greenberger1989}. Such states provide an
important resource for applications ranging from quantum
metrology~\cite{Pezze2018} to quantum error correction~\cite{Nielsen2011}.
However, these are among the most fragile many-body states because a single
error on any one of the $N$ qubits collapses the superposition, resulting in a
statistical mixture. Remarkably, despite their highly entangled nature, GHZ
states can be characterized by just two diagonal and two off-diagonal terms in
the $N$-particle density matrix. In contrast to quantifying the degree of
entanglement in general many-body states, which is extremely
challenging~\cite{Amico2008,Guhne2009,Islam2015}, the GHZ state fidelity
($\mathcal{F}>0.5$) constitutes an accessible witness for $N$-partite
entanglement~\cite{Sackett2000}. For these reasons, GHZ state creation can
serve as an important benchmark for characterizing the quality of any given
quantum hardware. Such states have been previously generated and characterized
by using systems of nuclear spins~\cite{Laflamme1998,Neumann2008}, individually
controlled optical photons~\cite{Bouwmeester1999,Pan2001,Wang2018a}, trapped
ions~\cite{Sackett2000,Leibfried2005,Monz2011,Friis2018}, and superconducting
quantum circuits~\cite{DiCarlo2010,Song2017a}. Large-scale superposition states
have also been generated in systems of microwave photons~\cite{Vlastakis2013}
and atomic ensembles without individual particle addressing~\cite{Pezze2018}.

\begin{figure}
    \centering
    \includegraphics[width=0.85\columnwidth]{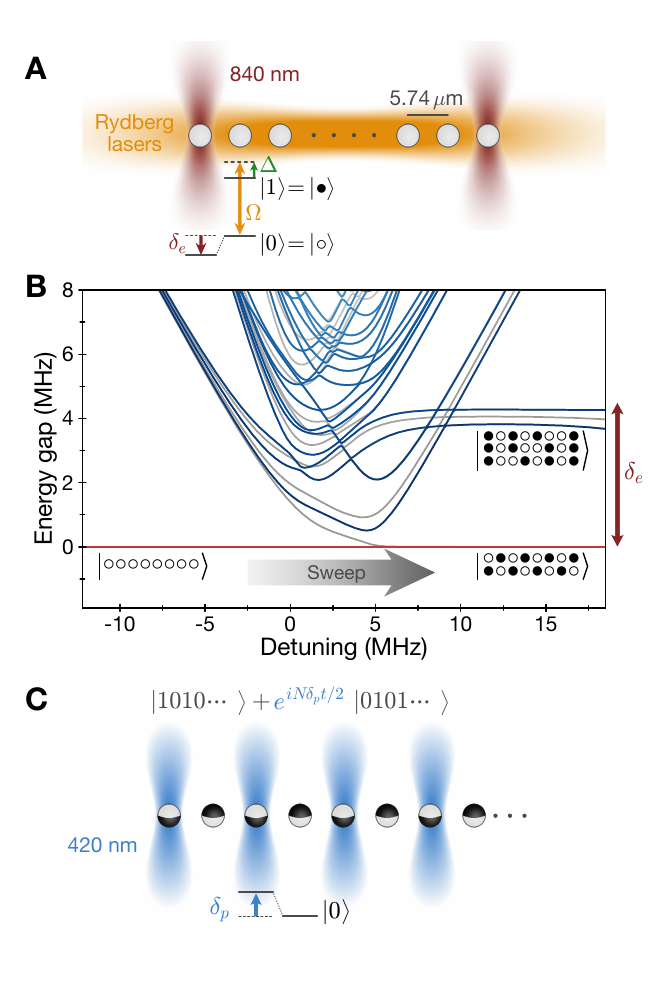}
    \caption{Experimental scheme and entanglement procedure.
        \textbf{A,}
        $^{87}{\rm Rb}$ atoms initially in a ground state
        $\left|0\right\rangle=\left|5S_{1/2},F=2,m_F=-2\right\rangle$ are
        coupled to a Rydberg state $\left|1\right\rangle=\left|70S_{1/2},
        m_J=-1/2\right\rangle$ by a light field with a coupling strength
        $\Omega/(2\pi)\leq 5\,$MHz and a variable detuning $\Delta$. Local
        addressing beams at $840\,$nm target the edge atoms, reducing the
        energy of $\left|0\right\rangle$ at those sites by a light shift
        $\delta_e$.
        \textbf{B,}
        Many-body energy gap spectrum of $N=8$ atoms, including energy shifts
        on the edge atoms. For positive detuning, the states with one ground
        state atom on the edges are favored over states with a Rydberg atom on
        both edges. An adiabatic pathway connects the state
        $\left|G_N\right\rangle=\left|000\cdots\right\rangle$ with the two GHZ
        components. Gray lines in the spectrum are energies associated with
        antisymmetric states, which are not coupled to the initial state by
        Hamiltonian~(2). 
        \textbf{C,}
        Method to control the phase $\phi$ of GHZ states. Every other site of
        the array is illuminated with a local addressing beam at $420\,$nm,
        which imposes a negative differential light shift $\delta_p$ on the
        $\left|0\right\rangle$-to-$\left|1\right\rangle$ transition. The offset
        in state $\left|0101\cdots\right\rangle$ relative to
        $\left|1010\cdots\right\rangle$ leads to an evolving dynamical phase.
        }
    \label{fig:Fig1}
\end{figure}

Here, we demonstrate the preparation of $N$-particle GHZ states
\begin{equation}
    \left|{\rm GHZ}_N\right\rangle =
    \frac{1}{\sqrt{2}}\left(\left|0101\cdots\right\rangle + \left|1010\cdots\right\rangle\right)
    \label{eq:GHZState}
\end{equation}
in a one dimensional array of individually trapped neutral $^{87}{\rm Rb}$
atoms, in which the qubits are encoded in an atomic ground state
$\left|0\right\rangle$ and a Rydberg state $\left|1\right\rangle$ (phase
convention is provided in~\cite{SOM}). Our entangling operation relies on the
strong van-der-Waals interaction between atoms in states $\left|1\right\rangle$
and on engineering the energy spectrum of the quantum many-body system to allow
for a robust quantum evolution from an initial product state to a GHZ state.
For both generating and characterizing GHZ states (Fig.~\ref{fig:Fig1}), all
the atoms were homogeneously coupled to the Rydberg state
$\left|1\right\rangle$ by means of a two-photon transition with an effective
coupling strength $\Omega(t)$ and detuning
$\Delta(t)$~\cite{Labuhn2016,Bernien2017}. In addition, we used addressing beams
to introduce local energy shifts $\delta_i$ on specific sites $i$ along the
array (Fig.~\ref{fig:Fig1}A). The resulting many-body Hamiltonian is
\begin{equation}
    \frac{H}{\hbar} = \frac{\Omega(t)}{2}\sum_{i=1}^N \sigma_x^{(i)} - \sum_{i=1}^N \Delta_i(t)
        n_i + \sum_{i<j}\frac{V}{\left|i-j\right|^6}n_i n_j
    \label{eq:Hamiltonian}
\end{equation}
where $\sigma_x^{(i)}=\left|0\right\rangle\!\left\langle1\right|_i +
\left|1\right\rangle\!\left\langle 0\right|_i$ is the qubit flip operator,
$\Delta_i(t)=\Delta(t)+\delta_i$ is the local effective detuning set by the
Rydberg laser and the local light shift,
$n_i=\left|1\right\rangle\!\left\langle 1\right|_i$ is the number of Rydberg
excitations on site $i$, and $V$ is the interaction strength of two Rydberg
atoms on neighboring sites. The separation between adjacent sites was chosen
so that the nearest-neighbor interaction $V=2\pi\cdot24\,$MHz $\gg\Omega$
results in the Rydberg blockade~\cite{Jaksch2000,Wilk2010,Isenhower2010a},
forbidding the simultaneous excitation of adjacent atoms into the state
$\left|1\right\rangle$.

To prepare GHZ states, we uesd arrays with an even number $N$ of atoms. For
large negative detuning $\Delta$ of the Rydberg laser, the many-body ground
state of the Hamiltonian~\eqref{eq:Hamiltonian} is
$\left|G_N\right\rangle=\left|0000\cdots\right\rangle$. For large uniform
positive detuning $\Delta_i = \Delta$, the ground-state manifold consists of
$N/2+1$ nearly degenerate classical configurations with $N/2$ Rydberg
excitations. These include in particular the two target antiferromagnetic
configurations $\left|A_N\right\rangle = \left|0101\cdots01\right\rangle$ and
$\left|\overline{A}_N\right\rangle =
\left|1010\cdots10\right\rangle$~\cite{Islam2013}, as well as other states with
nearly identical energy (up to a weak second-nearest neighbor interaction),
with both edges excited, such as $\left|10010\cdots01\right\rangle$. To isolate
a coherent superposition of states $\left|A_N\right\rangle$ and
$\left|\overline{A}_N\right\rangle$, we introduced local light shifts $\delta_e$
using off-resonant laser beams at $840\,$nm, generated with an acousto-optic
deflector (AOD), which energetically penalize the excitation of edge atoms
(Fig.~\ref{fig:Fig1}A), and effectively eliminate the contribution of undesired
components. In this case, the ground state for positive detuning is given by
the GHZ state~\eqref{eq:GHZState} and there exists in principle an adiabatic
pathway that transforms the state $\left|G_N\right\rangle$ into $\left|{\rm
GHZ}_N\right\rangle$ by adiabatically increasing $\Delta(t)$ from negative to
positive values (Fig.~\ref{fig:Fig1}B).


\begin{figure*}
    \centering
    \includegraphics[width=1.4\columnwidth]{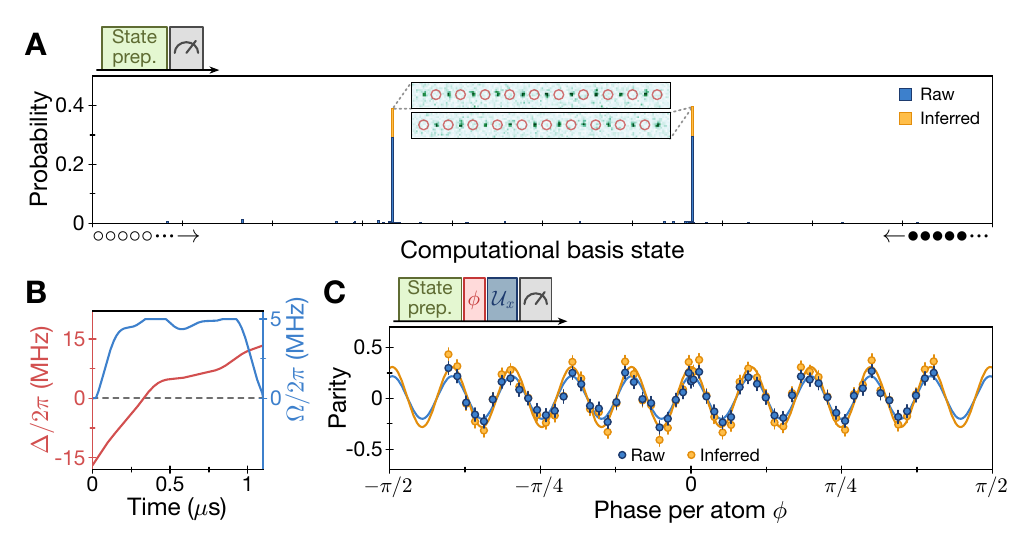}
    \caption{Characterization of a $20$-atom GHZ state.
\textbf{A,}
Probability of observing different patterns, showing a large population
of the two target patterns out of $2^{20}=1,048,576$ possible states.
Shown here are the raw measured values (blue bars) and the
populations inferred by using maximum likelihood estimation (orange bars)
for the two target states. (Insets) fluorescence images of the two
target patterns, where red circles mark empty sites corresponding to
atoms in state $\left|1\right\rangle$.
\textbf{B,}
Optimal control pulse used for state preparation.
\textbf{C,}
Parity oscillations produced by acquiring a relative phase between the
GHZ components. We apply a staggered field with a shift of
$\delta_p/(2\pi)=\pm3.8\,$MHz on all sites, followed by an operation
$\mathcal{U}_x$ so that subsequent parity measurements are sensitive
to $\phi$~\cite{SOM}. From the population measurement and the
oscillation amplitude, we infer a lower bound on the $20$-atom GHZ
fidelity of $\mathcal{F}\geq0.542(18)$. Error bars denote $68\%$
confidence intervals.
        }
    \label{fig:Fig2}
\end{figure*}
In practice, the time necessary to adiabatically prepare such a GHZ state grows
with system size and becomes prohibitively long for large $N$, owing to small
energy gaps in the many-body spectrum. To address this limitation, we used
optimal control methods to find laser pulses that maximize the GHZ state
preparation fidelity while minimizing the amount of time necessary. Our
specific implementation, the remote dressed chopped-random basis algorithm
(RedCRAB)~\cite{Rach2015,Heck2018}, yields optimal shapes of the laser
intensity and detuning for the given experimental conditions~\cite{SOM}. For
$N\leq8$ atoms, we performed this optimization using $\delta_e/(2\pi) \approx
-4.5\,$MHz light shifts on the edge atoms. For larger systems of $N>8$, the
preparation was found to be more robust by increasing the edge light shifts to
$\delta_e/(2\pi) \approx -6\,$MHz and adding $\delta_{4,N-3}/(2\pi) \approx
-1.5\,$MHz light shifts on the third site from both edges.

Our experiments are based on the optical tweezer platform and experimental
procedure described previously~\cite{Bernien2017}. After the initialization of
a defect-free $N$-atom array, the traps were switched off while the atoms were
illuminated with the Rydberg and local light shift beams. The internal state of
the atoms is subsequently measured by imaging state $\left|0\right\rangle$
atoms recaptured in the traps, while Rydberg atoms are repelled by the trapping
light~\cite{deLeseleuc2018}. The results of such experiments for a $20$-atom array
are demonstrated in Fig.~\ref{fig:Fig2}. After applying the optimized pulse
shown in Fig.~\ref{fig:Fig2}B, we measured the probability of observing
different patterns $p_n = \left\langle n\right|\rho\left|n\right\rangle$ in the
computational basis, where $\rho$ is the density operator of the prepared
state. The measured probability to observe each one of the $2^{20}$ possible patterns
in a $20$-atom array is shown in Fig.~\ref{fig:Fig2}A. The states
$\left|A_{20}\right\rangle$ and $\left|\overline{A}_{20}\right\rangle$ clearly
stand out (Fig.~\ref{fig:Fig2}A, blue bars) with a combined probability of
$0.585(14)$ and almost equal probability of observing each one. 

To characterize the experimentally prepared state $\rho$, we evaluated the GHZ
state fidelity 
\begin{equation}
    \mathcal{F} = \left\langle{\rm GHZ}_N\right|\rho\left|{\rm GHZ}_N\right\rangle
    = \frac{1}{2} \left( p_{A_N} + p_{\overline{A}_N}  + c_N + c_N^*\right)
\end{equation}
where $p_{A_N}$ and $p_{\overline{A}_N}$ are the populations in the target
components and $c_N =
\left\langle\overline{A}_N\right|\rho\left|A_N\right\rangle$ is the
off-diagonal matrix element, which can be measured by using the maximal
sensitivity of the GHZ state to a staggered magnetic field. Specifically,
evolving the system with the Hamiltonian $H_p = \hbar\delta_p/2\sum_{i=1}^N
(-1)^i \sigma_z^{(i)}$, the amplitude $c_N$ acquires a phase $\phi$ at a rate
of $\dot{\phi}=N\delta_p$. Measuring an observable that oscillates at this
frequency provides a lower bound on the coherence $\left|c_N\right|$ through
the oscillation contrast~\cite{Garttner2017,SOM}. In our experiments, the
staggered field was implemented by applying off-resonant focused beams of equal
intensity at $420\,$nm, generated by another AOD, to every other site of the
array (Fig.~\ref{fig:Fig1}C), resulting in a local energy shift
$\delta_p$~\cite{SOM}. Subsequently, we drove the atoms resonantly, applying a
unitary operation $\mathcal{U}_x$ in order to change the measurement
basis~\cite{SOM}, so that a measurement of the parity $\mathcal{P}=\prod_i
\sigma_z^{(i)}$ becomes sensitive to the phase of $c_N$. The measured parity is
shown in Fig.~\ref{fig:Fig2}C as a function of the phase accumulated on each
atom, demonstrating the coherence of the created state.

To extract the entanglement fidelity for large atomic states, we carefully
characterized our detection process used to identify atoms in
$\left|0\right\rangle$ and $\left|1\right\rangle$ because it has a small but
finite error. We have independently determined the probability to misidentify
the state of a particle to be $p(1|0)=0.0063(1)$, and
$p(0|1)=0.0227(42)$~\cite{SOM}. Subsequently, we use a maximum-likelihood
estimation procedure to infer the properties of created states on the basis of
the raw measurement results. Using this procedure, we infer a probability of
preparing states $\left|A_{20}\right\rangle$ and
$\left|\overline{A}_{20}\right\rangle$ to be $0.782(32)$ (Fig.~\ref{fig:Fig2}A,
orange bars) and an amplitude of oscillation of $0.301(18)$
(Fig.~\ref{fig:Fig2}C, orange points). From these measurements, we extracted a
lower bound for the $20$-atom GHZ state fidelity of $\mathcal{F}\geq0.542(18)$.

\begin{figure*}
    \centering
    \includegraphics[width=1.2\columnwidth]{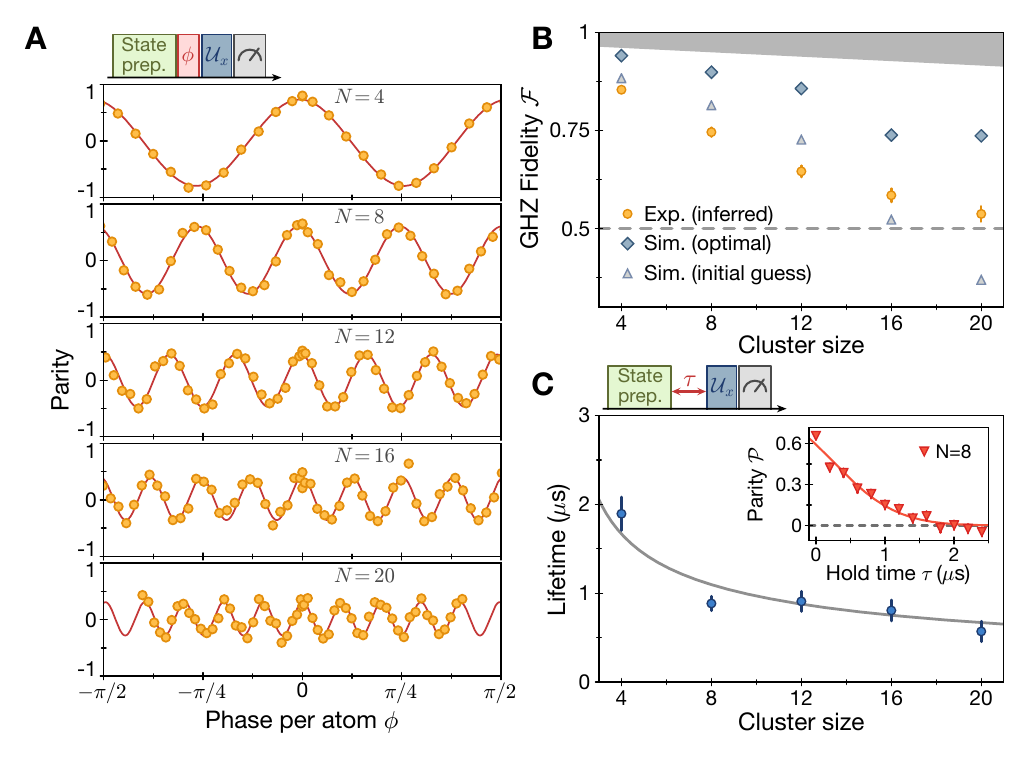}
    \caption{Quantifying entanglement for different system sizes.
        \textbf{A,}
        Parity oscillations measured on different system sizes. We apply a
        staggered field with a shift of $\delta_p/(2\pi)=\pm3.8\,$MHz on all
        sites and observe a scaling of the phase accumulation rate proportional
        to the system size $N$.
        \textbf{B},
        Inferred GHZ fidelity for different system sizes (orange
        circles)~\cite{SOM}. Blue diamonds show the result of simulations that
        account for dephasing during state preparation, decay from off-resonant
        photon scattering, and imperfect detection of coherence through parity
        oscillations~\cite{SOM}. Pale blue triangles show identical simulations
        for the initial guess pulses for the RedCRAB optimization, consisting
        of a $T=1.1\,\mu$s linear detuning sweep and
        $\Omega(t) = \Omega_{\rm max}[1 - \cos^{12}(\pi t/T)]$. The gray shaded
        area marks a region not measurable with our parity observable~\cite{SOM}.
        \textbf{C},
        Lifetime of the GHZ state coherence. For all system sizes $N$, we
        measure the state parity after a variable delay following the GHZ state
        preparation, which (inset) decays to zero. We fit the individual
        parity data to the tail of a Gaussian decay curve because we assume
        that the dephasing started during state preparation - before
        $\tau=0$.  The gray line shows a theoretical prediction with no free
        parameters, accounting for known dephasing mechanisms in our system.
    }
    \label{fig:Fig3}
\end{figure*}

This protocol was applied for multiple system sizes of $4\leq N \leq20$, using
$1.1\,\mu$s control pulses optimized for each $N$ individually. Consistent with
expected GHZ dynamics~(Fig.~\ref{fig:Fig1}C)~\cite{Leibfried2005}, the
frequency of the measured parity oscillations grows linearly with $N$
(Fig.~\ref{fig:Fig3}A). Extracting the GHZ fidelity from these measurements
shows that we surpass the threshold of $\mathcal{F}=0.5$ for all system sizes
studied (Fig.~\ref{fig:Fig3}B and table~\ref{table:MeasuredFidelities}). We
further characterized the lifetime of the created GHZ state by measuring the
parity signal after a variable delay (Fig.~\ref{fig:Fig3}C).  These
observations are most consistent with Gaussian decay, while characteristic
lifetimes are reduced relatively slowly for increasing system sizes, indicating
the presence of a non-Markovian environment~\cite{Nielsen2011,Monz2011}. 

\begin{figure}
    \centering
    \includegraphics[width=1.03\columnwidth]{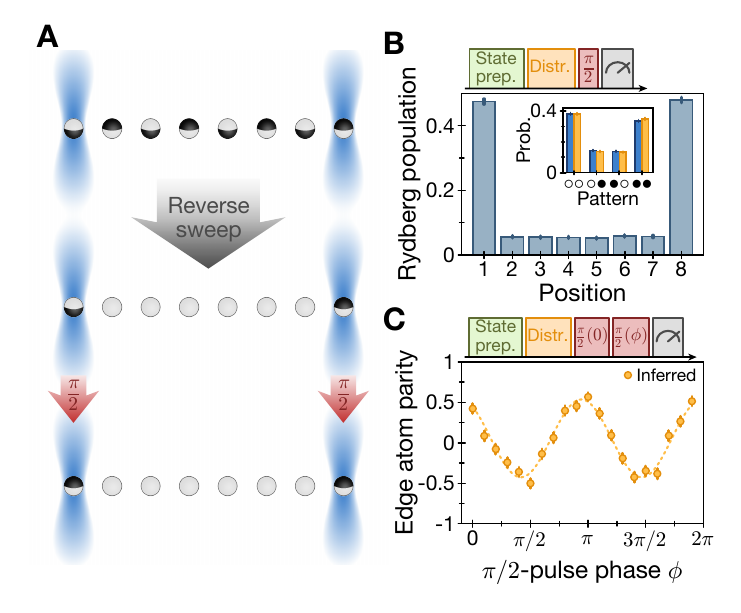}
    \caption{Demonstration of entanglement distribution.
        \textbf{A,}
        Experimental protocol for $N=8$. Edge atoms are addressed by light
        shift beams, and a reverse sweep of the Rydberg laser detuning is
        performed to disentangle the bulk of the array, leaving a Bell state
        $\left|\Psi^+\right\rangle\propto\left|1\cdots0\right\rangle+\left|0\cdots1\right\rangle$
        on the edge. A $\pi/2$ pulse resonant only with the edge atoms is
        applied to convert the state $\left|\Psi^+\right\rangle$ to
        $\left|\Phi^+\right\rangle\propto\left|0\cdots0\right\rangle+\left|1\cdots1\right\rangle$.
        \textbf{B,}
        Measured Rydberg populations on each site after entanglement
        distribution. (Inset) Probabilities for different patterns on the edge
        atoms, which are consistent with the Bell state
        $\left|\Phi^+\right\rangle$. Blue bars show the raw data, while orange
        bars are the statistically inferred probabilities given our detection
        errors.
        \textbf{C,}
        Measurement of the Bell state coherence. GHZ entanglement is
        distributed to the edges, and a $\pi/2$ pulse is applied at laser phase
        $\phi=0$, followed by a second $\pi/2$ pulse at varying phase $\phi$.
        The amplitude of the parity oscillation provides a lower bound on the
        coherence of the Bell state, yielding a fidelity of
        $\mathcal{F}\geq0.605(13)$.
        }
    \label{fig:Fig4}
\end{figure}

As an application of our entanglement manipulation technique, we demonstrate
its use for entanglement distribution between distant atoms. Specifically, we
consider the preparation of Bell states between atoms at the two opposite edges
of the array. Our approach was based on first creating the GHZ state by using
the above procedure, followed by an operation that disentangles all but two
target atoms. The latter is realized by shifting the transition frequencies of
the two target edge atoms by using two strong, blue-detuned addressing beams at
$420\,$nm. Subsequently, we performed a reverse detuning sweep of the Rydberg
laser that effectively disentangles all atoms except those at the edges. The
resulting state corresponds to a coherent superposition of two pinned
excitations that can be converted into a Bell state $\left|\Phi^+\right\rangle
= (\left|00\right\rangle + \left|11\right\rangle)/\sqrt{2}$ by applying a
resonant $\pi/2$ pulse on the edge atoms (Fig.~\ref{fig:Fig4}A).

To demonstrate this protocol experimentally, we prepare a GHZ-state of 8 atoms,
and turn on the detuned $420\,$nm addressing beams on the edge atoms, resulting
in a shift of $\delta_{1,8}/(2\pi) = 6\,$MHz. We then used an optimized Rydberg
laser pulse to distribute the entanglement and observed the patterns
$\left|00000000\right\rangle$ and $\left|10000001\right\rangle$ with a total
probability of $0.729(9)$ after accounting for detection errors
(Fig.~\ref{fig:Fig4}B). We verified the coherence of the remote Bell pair by
applying an additional $\pi/2$ pulse with a variable laser phase, and observed
parity oscillations with an amplitude of $0.481(24)$ (Fig.~\ref{fig:Fig4}C).
Combining these results, we obtained the edge atom Bell state fidelity of
$0.605(13)$.

Regarding our experimental observations, the optimal control provides a
substantial improvement over na\"ive analytic pulses (Fig.~\ref{fig:Fig3}B),
while bringing our protocol close to the speed set by a more conventional
protocol of building up entanglement through a series of two-qubit
operations~\cite{SOM}. By contrast, a simple linear detuning sweep only allows
for the creation of GHZ states for $N\leq16$ within a fixed $1.1\,\mu$s window
(Fig.~\ref{fig:Fig3}B), even under ideal conditions. Our analysis reveals that
the reason for this improvement stems from diabatic excitations and
de-excitations in the many-body spectrum, related to the recently proposed
mechanisms for quantum optimization speedup~\cite{Farhi2014,Zhou2018,SOM}.

The measured entanglement fidelity is partially limited by imperfect qubit
rotations used for parity measurements. Specifically, the qubit rotation
operation $\mathcal{U}_x$ in our experiment is induced by an interacting
Hamiltonian, which complicates this step~\cite{SOM}. The resulting evolution
can be understood in terms of quantum many-body
scars~\cite{Bernien2017,Turner2018a}, which gives rise to coherent qubit
rotations, even in the presence of strong interactions. The deviations from an
ideal parity measurement arise from the Rydberg blockade constraint and
long-range interactions~\cite{SOM}. These grow with the system size, resulting
in finite fidelities even for a perfect initial GHZ state
(Fig.~\ref{fig:Fig3}B, gray shaded area). Our quoted fidelity values do not
include the correction for this imperfection and represent the lower bound on
the actual GHZ state fidelities.

Entanglement generation, manipulation and lifetime are further limited by
several sources of decoherence. The finite temperature of the atoms leads to
random Doppler shifts on every site as well as position fluctuations that
influence interaction energies. These thermal dephasing mechanisms lead to a
Gaussian decay of the GHZ state coherence, whose time scale decreases with the
system size as $1/\sqrt{N}$, which is in good agreement with our observations
(Fig.~\ref{fig:Fig3}C). Additionally, off-resonant laser scattering introduces
a small rate of decoherence on each site in the array. We find that numerical
simulations of the state preparation accounting for these imperfections predict
higher GHZ fidelities than those obtained experimentally (Fig.~\ref{fig:Fig3}B)
\cite{SOM}. We can attribute this discrepancy to several additional
sources of errors. Laser phase noise likely contributes to the finite fidelity
of the state preparation. Drifts in the beam positions of the Rydberg lasers
can lead to changing light shifts, giving rise to uncontrolled detunings, and
drifts in the addressing beam positions can lead to an imbalance in the local
energy shifts and thereby in the populations of the two GHZ components,
limiting the maximum possible coherence. This analysis highlights the utility
of GHZ states for uncovering sources of errors. We emphasize that all of these
known error sources can be mitigated through technical improvements~\cite{SOM}.

Our experiments demonstrate a new promising approach for the deterministic
creation and manipulation of large-scale entangled states, enabling the
realization of GHZ-type entanglement in system sizes of up to $N=20$ atoms.
These results show the utility of this approach for benchmarking quantum
hardware, demonstrating that Rydberg atom arrays constitute a competitive
platform for quantum information science and engineering. Specifically, the
entanglement generation and distribution could potentially be used for
applications that range from quantum metrology and quantum networking to
quantum error correction and quantum computation. Our method can be extended by
mapping the Rydberg qubit states used here to ground-state hyperfine sublevels,
so that the entangled atoms can remain trapped and maintain their quantum
coherence over very long times~\cite{Wilk2010,Isenhower2010a,Picken2018,SOM}.
This could enable the sophisticated manipulation of entanglement and
realization of deep quantum circuits for applications such as quantum
optimization~\cite{Farhi2014,Zhou2018}.

During the completion of this work, we became aware of related results
demonstrating  large GHZ state preparation using superconducting quantum
circuits~\cite{Song2019,Wei2019}.

\section*{Acknowledgments}

We thank Dries Sels and Christian Reimer for helpful discussions. The authors
acknowledge financial support from the Center for Ultracold Atoms, the National
Science Foundation, Vannevar Bush Faculty Fellowship, the US Department of
Energy and the Office of Naval Research. H.L.  acknowledges support from the
National Defense Science and Engineering Graduate (NDSEG) fellowship. G.S.
acknowledges support from a fellowship from the Max Planck/Harvard Research
Center for Quantum Optics. J.C., S.M., and T.C. acknowledge funding from the EC
H2020 grants 765267 (QuSCo), 817482 (PASQuANS), and QuantERA QTFLAG; the DFG
SPP 1929 (GiRyd) and TWITTER; the IQST Alliance; and the Italian PRIN 2017.

\FloatBarrier

\bibliographystyle{Science}


\clearpage
\newpage

\renewcommand{\thefigure}{S\arabic{figure}}
\renewcommand{\thetable}{S\arabic{table}}
\setcounter{figure}{0}
\setcounter{equation}{0}

{
\renewcommand{\addcontentsline}[3]{}
\section{Supplementary Materials}
}

\subsection{Experimental setup}

The Rydberg excitations are enabled by a two-color laser system at $420\,$nm
and $1013\,$nm wavelength. The $420\,$nm light is derived from a
frequency-doubled titanium sapphire laser (M Squared SolsTiS 4000 PSX F) locked
to an ultrastable reference cavity (by Stable Laser Systems).  The $1013\,$nm
light is obtained from a high-power fiber amplifier (ALS-IR-1015-10-A-SP by
Azur Light Systems). The seed light is derived from a Fabry-P\'erot laser diode
injection locked to an external cavity diode laser (CEL002 by MOGLabs)
stabilized to the same reference cavity and filtered by the cavity
transmission~\cite{Levine2018}.  The detuning of both Rydberg lasers to the
intermediate state $\left|6P_{3/2}, F=3, m_F=-3\right\rangle$ is approximately
$2\pi\times 2\,$GHz. The individual Rabi frequencies of the two Rydberg lasers
are $\Omega_{\rm 420}/(2\pi) \approx 174\,$MHz and $\Omega_{1013}/(2\pi)
\approx 115\,$MHz. This gives a two-photon Rabi frequency of $\Omega =
\Omega_{420} \Omega_{1013}/(2\Delta) \approx 2\pi\times5\,$MHz. We define the
local phases of each atom's states $\left|0\right\rangle$ and
$\left|1\right\rangle$ in the reference frame associated with the local phases
of Rydberg excitation lasers, such that the two GHZ components have a relative
phase $\phi=0$ after state preparation.

To drive the optimal control pulses, we modulate the $420\,$nm Rydberg laser
with an acousto-optic modulator (AOM) driven by an arbitrary waveform generator
(AWG, M4i.6631-x8 by Spectrum). We correct the nonlinear response of the AOM to
the drive amplitude by a feed-forward approach to obtain the target output
intensity pattern. Furthermore, the AOM efficiency changes with changing
frequency, which we compensate by feeding forward onto the waveform amplitude
to suppress the intensity variations with frequency. In addition, the light
shift on the Rydberg transition from the $420\,$nm laser can be as large as
$2\pi\times4\,$MHz.  While the pulse intensity changes, this light shift
changes, modifying the detuning profile. We therefore correct the frequency
profile as a function of the pulse intensity to compensate this shift. These
steps ensure that the experimentally applied pulse is a faithful representation
of the desired profile.

The local addressing beam patterns are generated by two AODs (DTSX400-800 by AA
Opto-electronic), each driven by multiple frequencies obtained from an
arbitrary waveform generator (M4i.6631-x8 by Spectrum). 

\subsection{Optimal control}

Optimal control was originally developed as a tool to harness chemical
reactions to obtain the largest amount of desired products with given
resources, and then introduced in quantum information processing as a standard
way of designing quantum protocols and quantum
devices~\cite{Montangero2007,Brif2010,Cui2015,Koch2016} as well as in
manipulating quantum many-body systems to exploit complex
phenomena~\cite{Doria2011,Glaser2015,Brouzos2015,Lloyd2014,vanFrank2016,Muller2013,Caneva2009,Caneva2011a,Cui2017,Heck2018}.
Quantum optimal control theory identifies the optimal shape of a time-dependent
control pulse to drive a quantum many-body system to accomplish given task,
e.g. state preparation or quantum gate implementation. The quality of the
transformation is certified by a Figure of Merit (FoM) that can be calculated
or measured, e.g. the fidelity of the final state with respect to the target
one, the final occupation, or the energy.

\begin{figure}
    \centering
    \includegraphics[width=0.8\columnwidth]{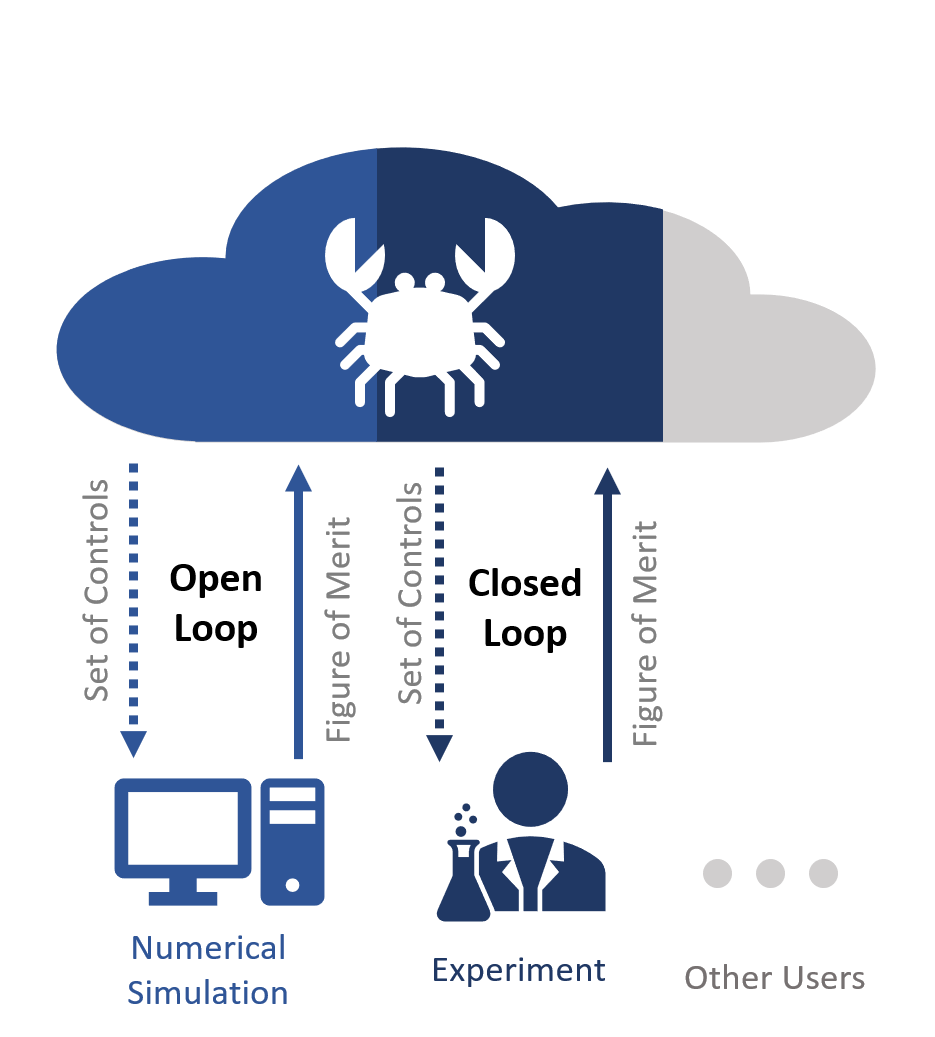}
    \caption{RedCRAB optimization loop. The remote dCRAB server generates and
        transmits a trial set of controls to the user, who evaluates the
        corresponding performance in terms of a FoM and sends the feedback
        information to the server, concluding one iteration loop. In the next
        loop, the server tends to generate an improved set of controls based on
        previous feedback information. The optimization continues until it
        converges. The FoM evaluation can be achieved either by numerical
        calculation (open-loop optimization) or experimental measurement
        (closed-loop optimization).}
    \label{fig:FigureSI1}
\end{figure}

\begin{figure*}
    \includegraphics[width=1.6\columnwidth]{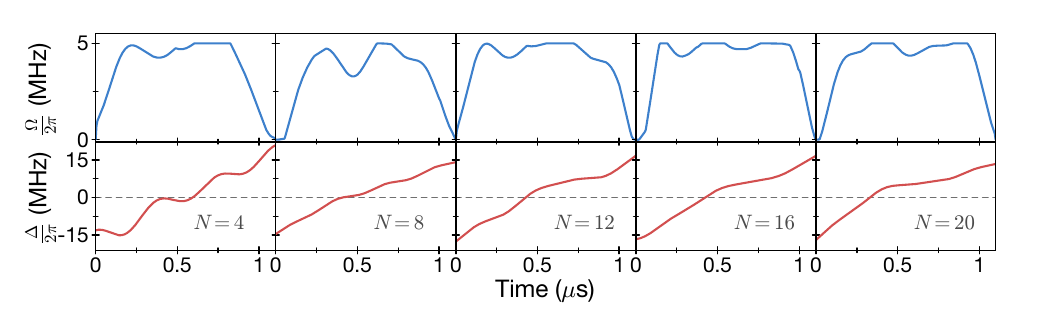}
    \caption{Optimal control pulse diagrams. Shown are the Rabi frequency (top)
    and detuning profiles (bottom) for the different system sizes investigated
    here.}
    \label{fig:FigureSI2}
\end{figure*}

In this work, the optimization is achieved through RedCRAB, the remote version
of the dressed Chopped RAndom Basis (dCRAB) optimal control via a cloud
server~\cite{Doria2011,Caneva2011a,Heck2018}. Within the optimization, control
fields such as the Rabi coupling $\Omega(t)$ are adjusted as
$\Omega(t)=\Omega_0(t) + f(t)$, where $\Omega_0(t)$ is an initial guess
function obtained from physical intuition or existing suboptimal solutions. The
correcting function $f(t)$ is expanded by randomized basis functions. In this
work, we chose a truncated Fourier basis. Thus, $f(t) = \Gamma(t)
\sum_{k=1}^{n_c}[A_k \sin(\omega_kt)+B_k \cos(\omega_kt)]$, where
$\omega_k=2\pi(k+r_k)/\tau$ are randomized Fourier frequencies with ${r_k}
\in[-0.5,0.5]$, $\tau$ is the final time, and $\Gamma(t)$ is a fixed scaling
function to keep the values at initial and final times unchanged, i.e.,
$\Gamma(0)=\Gamma(\tau)=0$. The optimization task is then translated into a
search for the optimal combination of $\{ A_k,B_k\}$ with a given $r_k$ to
maximize the fidelity between the target state and the time evolved state at
$\tau$. It can be solved by iteratively updating $\{ A_k,B_k\}$ using a
standard Nelder-Mead algorithm~\cite{Nelder1965}. In the basic version of the
CRAB algorithm, all $r_k$ are fixed and the local control landscape is explored
for all $n_c$ frequencies simultaneously. This leads to a restriction in the
number of frequencies that can be efficiently optimised. Using the dressed CRAB
(dCRAB) algorithm, only one Fourier frequency $\omega_k$ is optimised at a
time. We then move on to $\omega_{k+1}$ after a certain number of iterations of
the CRAB routine. This enables the method to include an arbitrarily large
number of Fourier components and deriving the solutions without -- whenever no
other constraints are present -- being trapped by local optima~\cite{Rach2015}.

In the RedCRAB optimization, the server generates and transmits a trial set of
controls to the client user, who will then evaluate the corresponding FoM and
communicates the feedback information to the server finishing one iteration
loop (Fig.~\ref{fig:FigureSI1}). The optimization continues
iteratively and the optimal set of controls, as well as the corresponding FoM
are derived. In the RedCRAB optimization, the user can either evaluate the FoM
by numerical calculation, namely open-loop optimization, or by experimental
measurement, which is called closed-loop optimization. In this work, open-loop
optimization was carried out only. The resulting controls could later serve as
the initial guess for a future closed-loop optimization. This last step would
ensure that the resulting controls are robust, since all unknown or not
modelled experimental defects and perturbations would automatically be
corrected for.

For the open-loop optimization of the pulse, we constrained the preparation
time to $1.1\,\mu$s and allowed the detuning $\Delta/(2\pi)$ to vary between
$-20\,$MHz and $20\,$MHz, while $\Omega/(2\pi)$ could vary between $0-5\,$MHz.
The resulting pulses are shown in Fig.~\ref{fig:FigureSI2}. While shorter
pulses can work sufficiently well for smaller system sizes, we use an equal
pulse duration for all $N$ for better comparability. We find that the optimized
pulses for larger systems appear smoother than for smaller system sizes, where
the pulses bear less resemblance to an adiabatic protocol. However, the
adiabaticity does not improve for larger system sizes, owing to the shrinking
energy gaps.

\subsection{Optimal control dynamics}

To gain insight into the timescales required to prepare a GHZ state in our
setup, we can compare our optimal control protocol with a minimal quantum
circuit consisting of a series of two-qubit gates that would achieve the same
task. In this circuit, a Bell pair is created in the first layer $p=1$ in the
middle of the array using the Rydberg blockade, which for our maximal coupling
strength of $\Omega/(2\pi)=5\,$MHz takes $100\,$ns$/\sqrt{2}$. The entanglement
can be spread to the two atoms adjacent to this Bell pair by simultaneously
applying  a pair of local $\pi$ pulses of $100\,$ns to those sites,
corresponding to controlled rotations. A sequence of such gate layers $
p=2,...,10$, including operations on qubit pairs and the free evolution of other
qubits, leads to the same GHZ state we prepare. This gate sequence requires
approximately $1\,\mu$s, which is within $10\%$ of the total evolution time
required in our optimal control sequence, which builds up the entanglement in
parallel. Furthermore, the fidelity of each layer of such a circuit effectively
acting on all $N=20$ qubits needs to be higher than $0.94$ to achieve the
20-qubit GHZ fidelity demonstrated in this work.

\begin{figure}
    \centering
    \includegraphics[width=0.95\columnwidth]{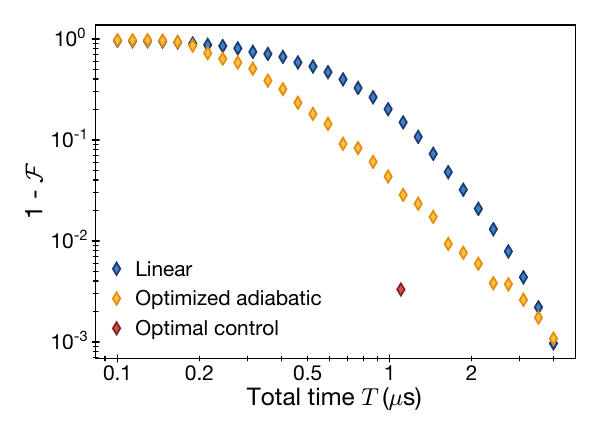}
    \caption{Comparison of ramping profile fidelities.
      Comparison of linear ramps (blue) to optimized adiabatic ramps (orange)
      for $N=12$ as a function of the total ramp time $T$. The optimal control
      pulse used in the experiment takes $T=1.1~\mu$s and achieves a higher
      fidelity than either the linear ramp or the optimized adiabatic ramp.
    }
    \label{fig:FigureSI3}
\end{figure}

It is interesting to compare this required evolution time with a parameter ramp
that tries to adiabatically connect the initial state to the GHZ state. To this
end, we parametrize the detuning and Rabi frequency as
$\Delta(s)=(1-s)\Delta_0+s\Delta_1$ and $\Omega(s)=\Omega_{\rm
max}[1-\cos^{12}(\pi s)]$ respectively. A na\"ive (unoptimized) linear ramp of
the detuning corresponds to choosing $s=t/T$. Alternatively, one can adjust the
local ramp speed to minimize diabatic transitions, for example by choosing
$s(t)$ minimizing 
$$
D=\left(\frac{ds}{dt}\right)^2\sum_{n>0}\frac{|\langle E_n(s)|\partial_sH(s)|E_0(s)\rangle|^2}{(E_n(s)-E_0(s))^2}
$$
during a ramp of duration $T$. Here $|E_n(s)\rangle$ are the instantaneous
eigenstates of the Hamiltonian $H(s)$ specified by the parameters $\Omega(s)$
and $\Delta(s)$, with $|E_0(s)\rangle$ denoting the instantaneous ground state.
In Fig.~\ref{fig:FigureSI3}, we show the results of numerical simulations
using both the linear sweep and a sweep that minimizes the strength of
diabatic processes quantified by $D$. Both sweep profiles require larger total
evolution time $T$ than the optimal control pulse to reach similar fidelities.

\begin{figure}
    \centering
    \includegraphics[width=\columnwidth]{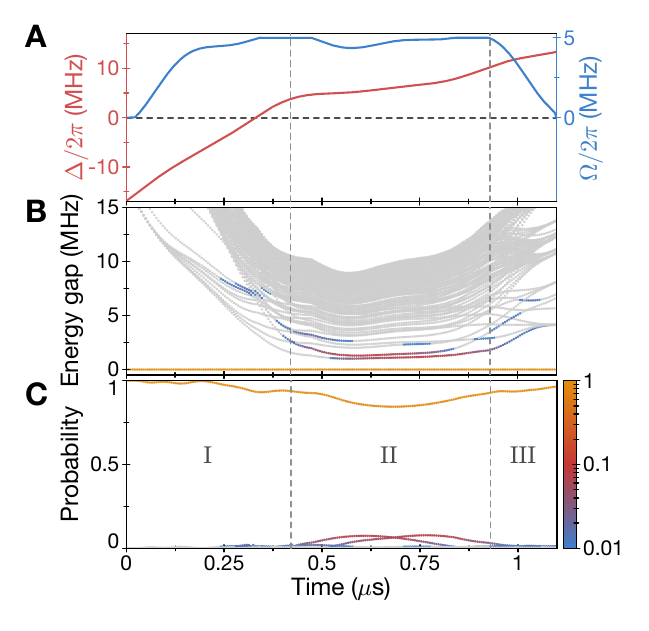}
    \caption{Dynamics of an optimized 20-atom GHZ state preparation.
        \textbf{A,}
        Optimized control parameters $\Omega(t)$ and $\Delta(t)$ for $N=20$ atoms.
        \textbf{B,}
        Energy eigenvalues of instantaneous eigenstates of the Hamiltonian
        relative to the ground state energy. The population in each energy
        eigenstate is color coded on a logarithmic scale. Light gray points
        correspond to populations smaller than $0.01$.
        \textbf{C,}
        Probability in each instantaneous eigenstate as the initial state
        evolves under the time-dependent Hamiltonian. The probability is
        dominated by the ground state and a few excited states. The time
        evolution is computed by exact numerical integration of Schr\"odinger’s
        equation, and 100 lowest energy eigenstates are obtained by using
        Krylov subspace method algorithms. For computational efficiency, we
        only consider the even parity sector of the Hamiltonian with no more
        than three nearest neighboring Rydberg excitations owing to the Rydberg
        blockade.
    }
    \label{fig:FigureSI4}
\end{figure}

To understand the origin of the speedup through optimal control, we numerically
simulate the corresponding evolution and analyze the population of the
instantaneous energy eigenstates (Fig.~\ref{fig:FigureSI4}). The optimal
control dynamics can be divided into three different regions: (I) A fast
initial quench, (II) a slow quench, and (III) a fast final quench. Even though
the change in the Hamiltonian parameters in region (I) is rather rapid, the
system remains mostly in the instantaneous ground state, with negligible
populations of the exited states, since the energy gap is large. In contrast,
in region (II) the parameters change slows down, reflecting the fact that the
energy gap becomes minimal. Unlike the adiabatic case however, one can observe
nontrivial population dynamics, with a temporary population of excited states.
Importantly, the optimal control finds a path in the parameter space such that
the population is mostly recaptured in the ground state at the end of region
(II). Finally, in region (III) the gap is large again and the system parameters
are quickly changed to correct also for higher order contributions. This
suggests that it actively uses diabatic transitions that go beyond the
adiabatic principle. This mechanism is related to the recently discussed
speedup in the context of the quantum approximate optimization algorithm
(QAOA)~\cite{Farhi2014,Zhou2018}.

\section{Quantifying detection}

The many-body dynamics involving coherent excitation to Rydberg states occurs
during a few-microsecond time window in which the optical tweezers are turned
off. After the coherent dynamics, the tweezers are turned back on, and atoms in
the ground state $\left|0\right\rangle$ are recaptured.  However, there is a
small but finite chance of losing these atoms. To quantify this error, we
perform the GHZ state preparation experiment while disabling the $420\,$nm
Rydberg pulse.  This keeps all atoms in state $\left|0\right\rangle$, and we
measure the loss probability to find a $0.9937(1)$ detection fidelity.

Atoms in state $\left|1\right\rangle$ on the other hand have a small chance of
being misidentified as being in state $\left|0\right\rangle$, as these
atoms can decay prematurely from the Rydberg state to the ground state and get
recaptured by the tweezers. This error probability can be measured by preparing
atoms at sufficiently large distances as to be non-interacting and applying a
calibrated $\pi$ pulse to transfer all atoms to $\left|1\right\rangle$ and
measure the probability of recapturing them. However, part of this signal is
given by the $\pi$ pulse infidelity, i.e. a small fraction of atoms which did
not get excited to $\left|1\right\rangle$ in the first place.

To quantify the $\pi$ pulse fidelity, we note that a Rydberg atom that decays
and is recaptured can decay either into the $F=2$ or $F=1$ ground states with
branching ratios $\alpha$ and $\beta$, respectively ($\alpha+\beta=1$).  Our
initial optical pumping of atoms into $\left|0\right\rangle$ has high fidelity
$>0.998$, measured using microwave spectroscopy on different sublevels of the
$F=2$ manifold. Thus, the final population of $F=1$ atoms should be given only
by Rydberg atom decay/recapture events. Following a $\pi$ pulse to excite
all atoms to the Rydberg state, the final measured population in $F=1$ is $p_1
= p \times \beta$, where $p$ is the total decay and recapture probability of a
Rydberg atom. Meanwhile, the final measured population in $F=2$ is $p_2 = p
\times \alpha + \epsilon$, which includes both decay events from Rydberg atoms
as well as residual population $\epsilon$ left from an imperfect $\pi$ pulse.
Experimentally, we separately measure the total recaptured ground state
population $(p_1+p_2)$, as well as the $F=1$ population $p_1$ only (by a
resonant push-out of $F=2$ atoms). We additionally can vary the overall
recapture probability $p$ by changing the depth of the tweezers that we
recapture atoms in, which changes the repulsive force exerted by the optical
tweezers on Rydberg atoms~\cite{deLeseleuc2018}. We measure $p_1$ and
$(p_1+p_2)$ at four different total recapture probabilities to extract the
$\pi$ pulse infidelity as $\epsilon = 0.006(3)$ (Fig.~\ref{fig:FigureSI5}).
From these measurements, we conclude a Rydberg detection fidelity of
$0.9773(42)$.

Detection errors of $\left|0\right\rangle$ can be mitigated by implementing
ground-state cooling in the tweezers~\cite{Kaufman2012a,Thompson2013a}, which
reduces the probability of loss after releasing the atoms. The detection
fidelity of $\left|1\right\rangle$ can be improved by using Rydberg states with
a longer radiative lifetime, actively ionizing the Rydberg atoms by electric or
optical fields, or by pulling them away from the trapping region with electric
field gradients.

\section{Accounting for detection imperfections}

The small imperfections in state detection of single qubits leads to a
prominent effect on the analysis of large systems. The probability for a single
detection error is sufficiently low that multiple errors per chain are very
unlikely, and we observe that the reduction in probability of observing the
correct GHZ pattern is dominated by these errors, as opposed to excitations of
the system (Fig.~\ref{fig:FigureSI6}A). This conclusion is further confirmed by
noting that near-ideal correlations extend across the entire system
(Fig.~\ref{fig:FigureSI7}).

To properly infer the obtained fidelities, we account for these 
imperfections using the following procedure:

\begin{figure}
    \centering
    \includegraphics[width=\columnwidth]{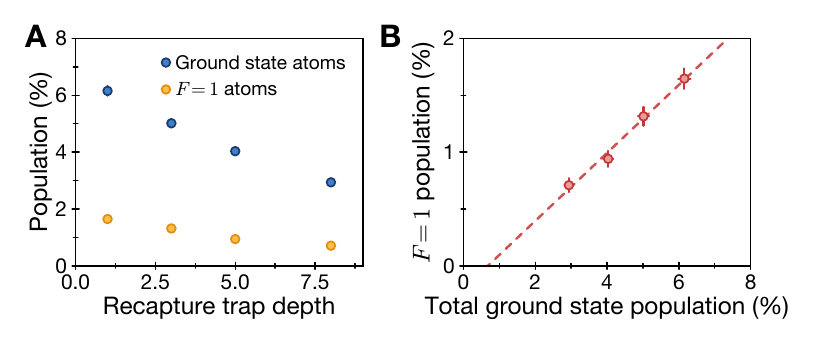}
    \caption{Quantifying detection errors.
        \textbf{A,}
        Measurement of the recaptured Rydberg atoms in the ground state (blue
        points) and in the $F=1$ ground-state manifold (orange points) as a
        function of the tweezer depth upon recapture.
        \textbf{B,}
        Recaptured populations in all ground state levels. The intersection
        with the horizontal axis gives an estimate of the atoms that were not
        excited to the Rydberg state, bounding the $\pi$ pulse fidelity.
    }
    \label{fig:FigureSI5}
\end{figure}

\begin{figure*}
    \centering
    \includegraphics[width=2\columnwidth]{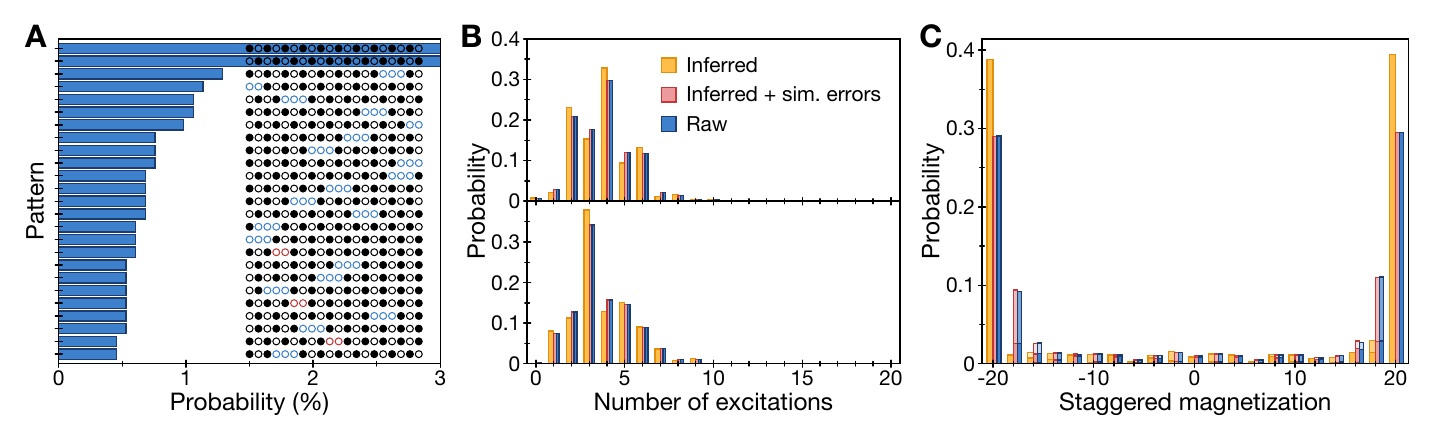}
    \caption{Inference of parity and populations.
        \textbf{A,}
        Histogram of observed patterns after preparing a 20-atom GHZ state.
        Open circles denote atoms in $\left|0\right\rangle$ and filled circles
        denote atoms in state $\left|1\right\rangle$. Blue domains mark regions
        where a single detection error has likely occurred, since such patterns
        are energetically costly at large positive detuning of the Rydberg
        laser. Red domains mark true domain walls, where the antiferromagnetic
        order is broken. Following the correct GHZ patterns, the 14 most
        observed patterns are consistent with a single detection error.
        \textbf{B,}
        Distribution of number of excitations measured for two different times
        of the parity oscillation for a 20-atom array, with the upper (lower)
        plot at $\phi=0$ ($\phi=\pi/20$) of phase accumulation per atom,
        showing a net positive (negative) parity. Blue bars show directly
        measured values, orange bars show the statistically inferred parent
        distribution, and red bars denote the parent distribution after adding
        simulated errors to compare to the raw data.
        \textbf{C,}
        Staggered magnetization $M_n$ extracted from the measurement of GHZ
        populations for 20 atoms. The vertically split bars with different
        shading denote different occurrences of number of excitations.
    }
    \label{fig:FigureSI6}
\end{figure*}

\textbf{Coherences:} The coherences are extracted from the amplitude of parity
oscillations. Each point in the parity oscillation is analyzed from the
measured distribution of the number of excitations in the system. We encode
this measured probability distribution in the vector $\mathbf{W}$, where $W_n$
is the probability to observe exactly $n$ excitations in the system ($0 \le n
\le N$). The true probability distribution of excitation numbers, prior to the
effect of detection errors, is denoted $\mathbf{V}$. Detection errors
transform this distribution according to a matrix $M$, where $M_{mn}$ encodes
the probability that a state with $n$ excitations will be detected as having
$m$ excitations. Each matrix element is calculated using combinatoric arguments
from the measured detection fidelities. We determine the true distribution
$\mathbf{V}$ as the one that minimizes the cost function $|M\mathbf{V} -
\mathbf{W}|^2$. (Fig.~\ref{fig:FigureSI6}B). This procedure is similar
to applying the inverse matrix $M^{-1}$ to the measured distribution
$\mathbf{W}$, but is more robust in the presence of statistical noise on the
measured distribution. Error bars on the inferred values are evaluated by
random sampling of detection fidelities, given our measured values and
uncertainties.

\textbf{Populations:} We carry out a similar procedure for the population data;
however, we are interested in assessing the probability of two particular
target states, which are defined not only by their number of excitations but
also by their staggered magnetizations $M_n = \sum_{i=1}^N (-1)^i \langle
\sigma_z^{(i)}\rangle $. Our procedure therefore operates by grouping all
possible microstates according to their common staggered magnetization and
number of excitations (Fig.~\ref{fig:FigureSI6}C). For $N$ particles, there are
in general $(N/2 + 1)^2$ such groups. As before, we denote the raw measured
distribution with respect to these groups as $\mathbf{W}$. We construct a
detection error matrix $M$ that redistributes populations between groups
according to the measured detection error rates. We optimize over all possible
true distributions to find the inferred distribution $\mathbf{V}$ that
minimizes the cost function $|M \mathbf{V} - \mathbf{W}|^2$. Following this
procedure, we sum the populations in the two groups that uniquely define the
two target GHZ components with a staggered magnetization of $\pm N$, and $N/2$
excitations.

\subsection{Bounding the GHZ state coherence}

We expand an experimental GHZ-like density matrix in the following
form
\begin{equation}
    \begin{aligned}
    \rho =& \alpha_1\left|A_N\right\rangle\!\left\langle A_N\right| + \alpha_2\left|\overline{A}_N\right\rangle\!\left\langle\overline{A}_N\right|\\
          &+ \left(\beta\left|A_N\right\rangle\!\left\langle\overline{A}_N\right| + \beta^*\left|\overline{A}_N\right\rangle\!\left\langle A_N\right|\right) + \rho'
    \end{aligned}
\end{equation}
where $\left|A_N\right\rangle=\left|0101\cdots\right\rangle$ and
$\left|\overline{A}_N\right\rangle=\left|1010\cdots\right\rangle$ are the
target GHZ components, $\alpha_i$ characterizes the diagonal populations in
these states $(0\leq\alpha_i\leq 1)$, $\beta$ characterizes the off-diagonal
coherence between these states $(0\leq|\beta|\leq 1/2)$, and $\rho'$ contains
all other parts of the density matrix.  The GHZ fidelity of state $\rho$ is
given by:
\begin{equation}
    \mathcal{F}= \left\langle{\rm GHZ}_N\right|\rho\left|{\rm GHZ}_N\right\rangle = \frac{\alpha_1 + \alpha_2}{2} + {\rm Re}(\beta)
\end{equation}

To measure the coherence $\left|\beta\right|$, we implement a staggered
magnetic field to which the target GHZ state is maximally sensitive:
\begin{equation}
  H_{\rm st} = \frac{\hbar\delta}{2}\sum_{i=1}^N (-1)^i \sigma_z^{(i)}
  \label{eq:StaggeredField}
\end{equation}

Applying $H_{\rm st}$ to the system for time $T$ results in unitary phase
accumulation $U(T) = \exp\left(-i H_{\rm st} T/\hbar\right)$. We then apply a
unitary $\mathcal{U}$ to the system and measure in the computational basis.
From repeated measurements, we calculate the expectation value of the global
parity operator $\mathcal{P} = \prod_i \sigma_z^{(i)}$ as a function of the
phase accumulation time $T$.  Denote the time-dependent expectation value
$E(T)$, where $-1 \le E(T) \le 1$.

We show that if $E(T)$ has a frequency component that oscillates at a frequency
of $N\delta$, then the amplitude of this frequency component sets a lower bound
for $|\beta|$.  Importantly, this holds for any unitary $\mathcal{U}$ used to
detect the phase accumulation.

\textbf{Proof:} The expectation value $E(T)$ can be written explicitly as the
expectation value of the time-evolved observable $\mathcal{P}\to U^\dagger(T)
\mathcal{U^\dagger P U} U(T)$. In particular,
\begin{equation}
    \begin{aligned}
        E(T) &= {\rm Tr}[\rho U^\dagger(T) \mathcal{U^\dagger P U} U(T)]\\
             &= \sum_n \left\langle  n\right|\rho U^\dagger(T) \mathcal{U^\dagger P U} U(T)\left|n\right\rangle
     \end{aligned}
    \label{eq:ExpectationValue}
\end{equation}
where $\left|n\right\rangle$ labels all computational basis states. Since the
phase accumulation Hamiltonian $H_{\rm st}$ is diagonal in the computational
basis, the basis states $\left|n\right\rangle$ are eigenvectors of $U(T)$ with
eigenvalues denoting the phase accumulation. Specifically,
\begin{equation}
  H_{\rm st}\left|n\right\rangle = \frac{\hbar\delta}{2}M_n\left|n\right\rangle
    \Rightarrow U(T)\left|n\right\rangle = e^{-i\delta T M_n/2}\left|n\right\rangle
\end{equation}
where $M_n$ is the staggered magnetization of state $\left|n\right\rangle$
defined earlier. The staggered magnetization of the state
$\left|A_N\right\rangle$ is maximal: $M_{A_N} = N$, and the staggered
magnetization of $\left|\overline{A}_N\right\rangle$ is minimal:
$M_{\overline{A}_N}=-N$.  Note that all other computational basis states have
strictly smaller staggered magnetizations. Inserting an identity operator in
Eq.~\eqref{eq:ExpectationValue}:

\begin{equation}
    \begin{aligned}
        E(T) =& \sum_{n,m} \left\langle n\right|\rho \left|m\right\rangle\!\left\langle m\right|U(T)^\dagger\mathcal{U^\dagger P U} U(T)\left|n\right\rangle\\
             =& \sum_{n,m} e^{-i\delta T (M_n - M_m)/2} \left\langle n\right|\rho\left|m\right\rangle\!\left\langle m\right|\mathcal{U^\dagger P U}\left|n\right\rangle
    \end{aligned}
\end{equation}

The highest frequency component comes from the states with maximally separated
staggered magnetization, $\left|n\right\rangle = \left|A_n\right\rangle$ and
$\left|m\right\rangle = \left|\overline{A}_n\right\rangle$. Separating out this
frequency component as $F(T)$, we obtain:
\begin{equation}
  \begin{aligned}
      F(T) =& 2{{\rm Re}}\left[e^{-iN\delta T}\left\langle A_N\right|\rho\left|\overline{A}_N\right\rangle\!\left\langle\overline{A}_N\right|\mathcal{U^\dagger P U}\left|A_N\right\rangle\right]\\
      =& 2{{\rm Re}}\left[\beta e^{-iN\delta T} \left\langle{\overline{A}_N}\right|\mathcal{U^\dagger P U}\left|A_N\right\rangle\right]
  \end{aligned}
\end{equation}

We note that the parity matrix element is bounded as $0 \le \left|\langle
\overline{A}_N | \mathcal{U^\dagger P U} |A_N\rangle \right| \le 1$.
Furthermore, the matrix element is real-valued and positive for the unitary
$\mathcal{U}$ considered in the experiment.  Fitting $F(T)$ to an oscillation
with amplitude $C \ge 0$ and phase $\phi$ according to $F(T) = C \cos (N\delta
T - \phi)$, we produce our lower bound for the off-diagonal coherence $\beta$:
\begin{equation}
  \left|\beta\right| \ge C/2; \quad \arg(\beta) = \phi
  \label{eq:inequality}
\end{equation}

\subsection{Parity detection}

The ideal observable to measure GHZ phase is the parity $\mathcal{P}_x =
\prod_i \sigma_x^{(i)}$. However, the presence of Rydberg interactions and the
Rydberg blockade prevents us from rotating all qubits such that we can measure
in this basis. Instead, in this work we generate a unitary
$\mathcal{U}_x=\exp\left(-i\Omega t/2\sum_i \sigma_x^{(i)} - iH_{\rm
int}t/\hbar\right)$ by resonantly driving all atoms in the presence of these
interactions given by $H_{\rm int}$ for a fixed, optimized time
(Fig.~\ref{fig:FigureSI8}), and subsequently measure the parity $\mathcal{P} =
\prod_i \sigma_z^{(i)}$ in the computational basis. The finite duration of the
unitary $\mathcal{U}_x$ incurs a small amount of additional infidelity, owing
both to dephasing and an additional laser scattering. However, we estimate that
this effect should only lead to small losses in fidelity on the percent level.

While it is not obvious that the parity observable used here is suitable, we
can understand the parity oscillations in the picture of weakly interacting
spin-$1$ particles defined on dimers of neighboring pairs of sites. For two
adjacent sites, we can define eigenstates of a spin-1 $S_z$ operator as
$\left|\circ\bullet\right\rangle = \left|-\right\rangle$,
$\left|\circ\circ\right\rangle = \left|0\right\rangle$, and
$\left|\bullet\circ\right\rangle = \left|+\right\rangle$. In this notation, the
antiferromagnetic  GHZ state we prepare is given by a ferromagnetic GHZ state
in the spin-1 basis:
\begin{equation}
    \left|{\rm GHZ}_N\right\rangle =
    \frac{1}{\sqrt{2}}\left(\left|+++\cdots\right\rangle +
    \left|---\cdots\right\rangle\right)
\end{equation}
We must express all operations on the GHZ state in this new notation. In
particular, the transverse field of the form $\hbar\Omega/2\sum_i \sigma_x^{(i)}$
applied to individual atoms gets transformed to an operation
$\hbar\Omega/\sqrt{2}\sum_j S_x^{(j)}$ on all dimers. Furthermore, the
staggered field $\hbar\delta/2 \sum_i (-1)^i \sigma_z^{(i)}$ we apply to
individual atoms to rotate the GHZ phase is equivalent to an operation of the
form $\hbar\delta \sum_j S_z^{(j)}$ acting on individual dimers.

\begin{figure}
    \centering
    \includegraphics[width=\columnwidth]{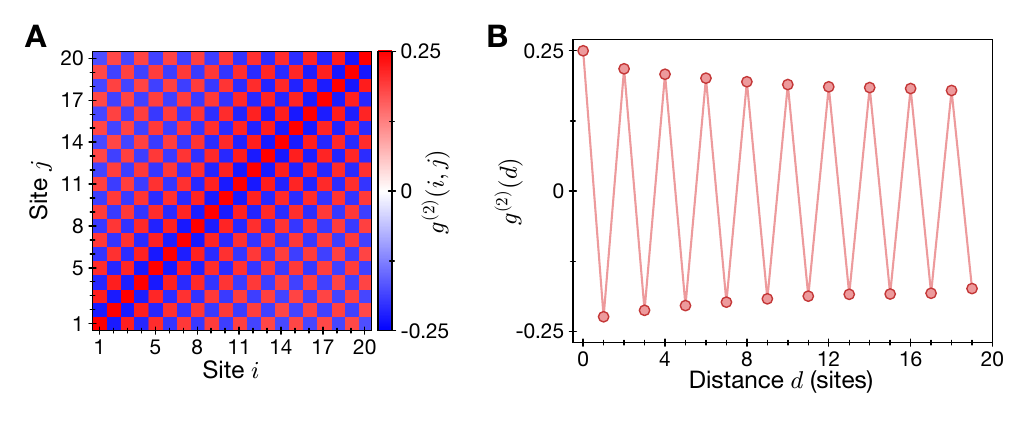}
    \caption{Density-density correlations for a $20$-atom GHZ state.
        \textbf{A},
        We evaluate the correlation function $g^{(2)}(i,j) = \left\langle n_i
        n_j\right\rangle - \left\langle n_i\right\rangle\left\langle
        n_j\right\rangle$ and observe strong correlations of Rydberg excitations
        across the entire system.
        \textbf{B},
        The density-density correlations over distance, given by $g^{(2)}(d) \propto
        \sum_i g^{(2)}(i,i+d)$ decay only very slowly throughout the array.}
    \label{fig:FigureSI7}
\end{figure}

The parity operator in the single-qubit basis $\mathcal{P}=\prod_i
\sigma_z^{(i)}$ can be transformed into the dimer basis as
\begin{equation}
    \mathcal{P} = \prod_j \left(-\left|+\right\rangle\!\left\langle +\right|_j-\left|-\right\rangle\!\left\langle -\right|_j + \left|0\right\rangle\!\left\langle 0\right|_j\right)
\end{equation}
by noting that the three dimer states are eigenstates of $\mathcal{P}$, i.e.
$\mathcal{P}\left|\pm\right\rangle = -\left|\pm\right\rangle$ and
$\mathcal{P}\left|0\right\rangle = \left|0\right\rangle$.

Assuming we begin from a GHZ state, applying a rotation on all dimers for a
duration given by $\Omega t = \pi/\sqrt{2}$ saturates the difference in
$\mathcal{P}$ between GHZ states of opposite phase. This shows that such a protocol
would be optimal if the dimer approximation were exact. However, interactions
between dimers cannot be neglected. In particular, the Rydberg blockade
suppresses configurations of the form $\left|\cdots-+\cdots\right\rangle$ owing
to the strong nearest-neighbor interaction $V$, and neighboring dimers of the
same type such as $\left|\cdots\pm\pm\cdots\right\rangle$ have a weak
interaction given by the next-to-nearest neighbor interaction strength
$V_2=V/2^6$. We can thus express the interactions in the system as

\begin{equation}
    \begin{aligned}
    \frac{H_{\rm int}}{\hbar} =& \sum_{j=1}^{N/2-1} V_2 \left|+\right\rangle\!\left\langle +\right|_j \left|+\right\rangle\!\left\langle +\right|_{j+1}
                    + V_2\left|-\right\rangle\!\left\langle-\right|_j \left|-\right\rangle\!\left\langle-\right|_{j+1}\\
                 &+ V \left|-\right\rangle\!\left\langle -\right|_j \left|+\right\rangle\!\left\langle +\right|_{j+1}
    \end{aligned}
    \label{eq:DimerInteractions}
\end{equation}
An exact simulation of the dimer rotation under the interaction
Hamiltonian~(11)
shows that both these interaction effects reduce the parity contrast by a small
amount. In the recently discussed context of quantum many-body
scars~\cite{Bernien2017,Turner2018a,Khemani2019,Ho2019a}, these effects of
residual interactions lead of small deviations from a stable periodic
trajectory through phase space.

\subsection{Staggered field calibration}

To apply the staggered field~(3),
we address each of the even sites in the array with a focused off-resonant
laser beam at $420\,$nm.  However, the unitary in question requires a staggered
field with opposite sign on every site. We compensate for the missing acquired
phase on the sites in between the addressed ones by shifting the phase of the
Rydberg laser, through a change in phase of the radio-frequency drive of the
AOM. The intensity of each addressing beam is measured by applying a spin-echo
sequence with an addressing pulse of variable duration to determine the light
shift on the Rydberg transition. We correct for inhomogeneous intensities so
that all atoms are subject to the same light shift.

\begin{figure}
    \centering
    \includegraphics[width=0.85\columnwidth]{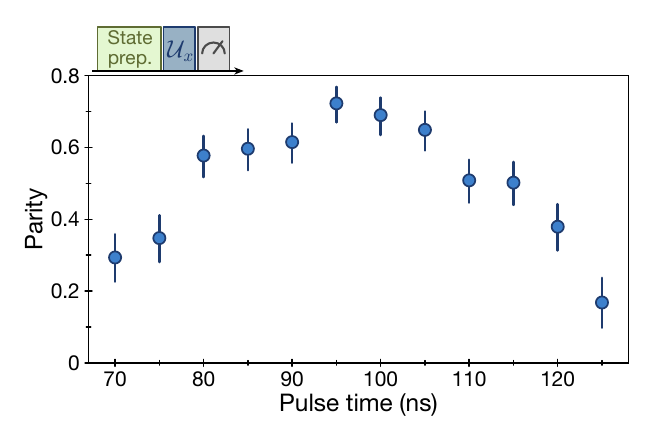}
    \caption{Parity signal measured as a function of the time the operation
    $\mathcal{U}_x$ is applied. The total time includes delays in the AOM response
    and the finite laser pulse rise time.}
    \label{fig:FigureSI8}
\end{figure}

\begin{figure}
    \centering
    \includegraphics[width=0.9\columnwidth]{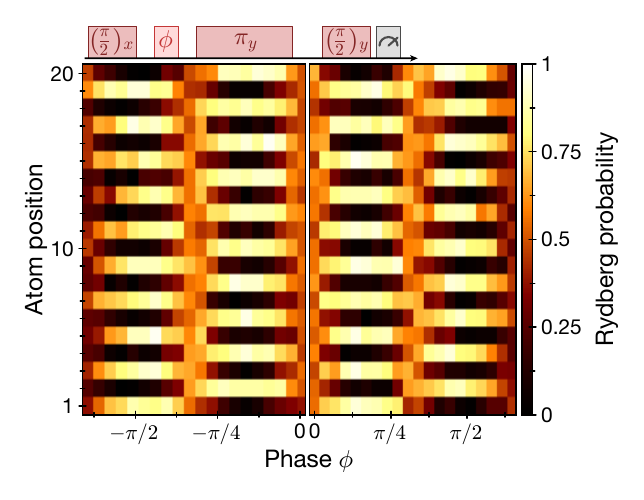}
    \caption{Phase accumulation measured on an array of 20 sites. The left
    panel demonstrates application of a negative staggered field by applying
    local addressing beams on the odd sites in the array. The right panel shows a
    positive staggered field by instead applying local addressing beams on the even
    sites in the array. Phase is accumulated on each site at a rate of $2\pi \times
    3.8~$MHz.}
    \label{fig:FigureSI9}
\end{figure}

We measure and calibrate the staggered field by measuring the effect of the
field on each atom individually. To do so, we alternately rearrange the atoms
to form different subsets of the $20$-atom system that are sufficiently far
apart to avoid interactions between them. In this configuration, every atom is
then subject to a $\pi/2$ rotation about the $x$-axis, followed by the
staggered field for variable duration, then a $\pi/2$ rotation about the
$y$-axis, to distinguish positive from negative phase evolution. With an
additional $\pi$ rotation about the $y$-axis, we perform a spin echo to
mitigate effects of dephasing. The outcome of this protocol is shown in
Fig.~\ref{fig:FigureSI9} and demonstrates the implementation of the staggered
magnetic field. By switching the local addressing beams to the opposite set of
alternating sites, we switch the sign of the staggered field, enabling the
measurement of both positive and negative phase accumulation.

\subsection{Measured GHZ fidelities}
For each system size $N$, we measure the GHZ populations and the GHZ coherence
by parity oscillations~(Figs.~\ref{fig:Fig2},~\ref{fig:Fig3} of the main text).
From the raw measurements, we infer the true GHZ fidelity using the maximum
likelihood procedure discussed in Section ``Accounting for detection
imperfections''.  All measured values are shown in table
\ref{table:MeasuredFidelities}.  Error bars on raw populations represent a
$68\%$ confidence interval for the measured value. Error bars on the raw
coherences are fit uncertainties from the parity oscillations. Error bars on
the inferred values include propagation of the uncertainty in the estimation of
the detection fidelities.

\begin{table*}[t]
\begin{center}
    \begin{tabular}{r | c | c | c | c | c}
  System size $N$ & 4 & 8 & 12 & 16 & 20\\
  \hline
  \hline
  Raw populations & 0.893(6) & 0.797(8) & 0.695(9) & 0.629(12) & 0.585(14) \\
  Inferred & 0.946(10) & 0.892(17) & 0.824(21) & 0.791(29) & 0.782(32) \\
  \hline
  Raw coherence & 0.710(12) & 0.516(11) & 0.371(10) & 0.282(11) & 0.211(11) \\
  Inferred & 0.759(11) & 0.598(16) & 0.462(19) & 0.373(19) & 0.301(18) \\
  \hline
  Raw fidelity & 0.801(7) & 0.657(7) & 0.533(7) & 0.455(8) & 0.398(9) \\
  Inferred & 0.852(7) & 0.745(12) & 0.643(14) & 0.582(17) & 0.542(18) \\
\end{tabular}
\end{center}
  \caption{Measured GHZ data for all system sizes. Errors denote $68\%$ confidence intervals.}
  \label{table:MeasuredFidelities}
\end{table*}

\subsection{Experimental Imperfections}
We identify a number of experimental imperfections that to varying degrees can
limit the coherent control of our atomic system.
\begin{enumerate}
  \item \textbf{Atomic temperature:} The atom temperature of $\sim 10\,\mu$K
      leads to fluctuating Doppler shifts in the addressing lasers of order
      $\sim 2\pi\times43~$kHz, as well as fluctuations in atomic position that
      leads to variation in Rydberg interactions strengths. These fluctuations
      are included in the simulations shown in the main text
      Fig.~\ref{fig:Fig3}. These effects can be dramatically reduced by
      improved atomic cooling, most notably by sideband cooling within the
      optical tweezers to the motional ground
      state~\cite{Kaufman2012a,Thompson2013a}.
  \item \textbf{Laser scattering:} The two-photon excitation scheme to our
      chosen Rydberg state leads to off-resonant scattering from the
      intermediate state, $6P_{3/2}$. This scattering rate has a timescale of
      $50-100 \mu$s for the two laser fields, and can be reduced by higher
      laser powers and further detuning from the intermediate state.
  \item \textbf{Rydberg state lifetime:} The $70S$ Rydberg state has an
      estimated lifetime of $150~\mu$s~\cite{Beterov2009a}, limited both by
      radiative decay and blackbody-stimulated transitions. This effect could
      be mitigated by selecting a higher Rydberg state with a longer lifetime
      or by cryogenic cooling of the blackbody environment.
\end{enumerate}

Additional error sources that may limit our coherence properties include laser
phase noise, which can be mitigated by better laser sources and stabilization
schemes, and fluctuations in local addressing beam intensities and positions,
which can be addressed by active feedback on the beam positions and improved
thermal and mechanical stability of the setup. Simulations predict that we
could go beyond the system sizes studied here. While GHZ states of $N=24$ could
be within reach with current parameters, generation of even larger GHZ states
should be feasible with the additional technical improvements discussed above.

\section{Ground-state qubit encoding}

The GHZ state parity could be more easily detected and manipulated if the
qubits were encoded in a basis of hyperfine sublevels of the electronic ground
state. In particular, one can consider two alternative qubit states
$|\tilde{0}\rangle=\left|5S_{1/2},F=1,m_F=-1\right\rangle$ and
$|\tilde{1}\rangle=\left|5S_{1/2},F=2,m_F=-2\right\rangle$. Rotations between
these states are possible through stimulated Raman transitions or microwave
driving, and the interactions are introduced by coupling $|\tilde{1}\rangle$ to
a Rydberg level $\left|r\right\rangle$. This type of hyperfine encoding has
been employed in multiple experiments with cold Rydberg
atoms~\cite{Wilk2010,Isenhower2010a,Picken2018}. To prepare GHZ states in this
basis, all atoms can be initialized in $|\tilde{1}\rangle$ and the system
transferred to the state
$|\tilde{1}r\tilde{1}r\cdots\rangle+|r\tilde{1}r\tilde{1}\cdots\rangle$ using
the method described in this work. A ground-state qubit $\pi$ pulse followed by
a $\pi$ pulse on the Rydberg transition transforms the state into
$|\tilde{0}\tilde{1}\tilde{0}\tilde{1}\cdots\rangle +
|\tilde{1}\tilde{0}\tilde{1}\tilde{0}\cdots\rangle$, enabling the long-lived
storage of entanglement. Additionally, local qubit rotations can flip the state
of every other site to prepare the canonical form of the GHZ state,
$|\tilde{0}\tilde{0}\tilde{0}\cdots\rangle +
|\tilde{1}\tilde{1}\tilde{1}\cdots\rangle$, which can achieve
entanglement-enhanced metrological sensitivity to homogeneous external fields
\cite{Pezze2018}. Incorporating this type of hyperfine qubit encoding with
Rydberg qubit control will be important for realizing quantum gates and deeper
quantum circuits in future experiments.


\begin{thebibliography}{10}
    \bibitem{Greenberger1989} D.~M. Greenberger, M.~A. Horne, A.~Zeilinger, {\it Bell's {{Theorem}}, {{Quantum Theory}} and {{Conceptions}} of the {{Universe}}\/}, M.~Kafatos, ed., Fundamental {{Theories}} of {{Physics}} ({Springer Netherlands}, Dordrecht, 1989), pp. 69--72.  
    \bibitem{Pezze2018} L.~Pezz\`e, A.~Smerzi, M.~K. Oberthaler, R.~Schmied, P.~Treutlein, {\it Rev.  Mod. Phys.\/} {\bf 90}, 035005 (2018).  
    \bibitem{Nielsen2011} M.~A. Nielsen, I.~L. Chuang, {\it Quantum {{Computation}} and {{Quantum Information}}: 10th {{Anniversary Edition}}\/} ({Cambridge University Press}, New York, NY, USA, 2011), 10th edn.
    \bibitem{Amico2008} L.~Amico, R.~Fazio, A.~Osterloh, V.~Vedral, {\it Rev. Mod. Phys.\/} {\bf 80}, 517 (2008).
    \bibitem{Guhne2009} O.~G\"uhne, G.~T\'oth, {\it Physics Reports\/} {\bf 474}, 1 (2009).  
    \bibitem{Islam2015} R.~Islam, {\it et~al.\/}, {\it Nature\/} {\bf 528}, 77 (2015).  
    \bibitem{Sackett2000} C.~A. Sackett, {\it et~al.\/}, {\it Nature\/} {\bf 404}, 256 (2000).  
    \bibitem{Laflamme1998} R.~Laflamme, E.~Knill, W.~H. Zurek, P.~Catasti, S.~V.~S. Mariappan, {\it Philosophical Transactions of the Royal Society of London. Series A: Mathematical, Physical and Engineering Sciences\/} {\bf 356}, 1941 (1998).  
    \bibitem{Neumann2008} P.~Neumann, {\it et~al.\/}, {\it Science\/} {\bf 320}, 1326 (2008).  
    \bibitem{Bouwmeester1999} D.~Bouwmeester, J.-W. Pan, M.~Daniell, H.~Weinfurter, A.~Zeilinger, {\it Phys.  Rev. Lett.\/} {\bf 82}, 1345 (1999).  
    \bibitem{Pan2001} J.-W. Pan, M.~Daniell, S.~Gasparoni, G.~Weihs, A.~Zeilinger, {\it Phys. Rev.  Lett.\/} {\bf 86}, 4435 (2001).  
    \bibitem{Wang2018a} X.-L. Wang, {\it et~al.\/}, {\it Phys. Rev. Lett.\/} {\bf 120}, 260502 (2018).  
    \bibitem{Leibfried2005} D.~Leibfried, {\it et~al.\/}, {\it Nature\/} {\bf 438}, 639 (2005).  
    \bibitem{Monz2011} T.~Monz, {\it et~al.\/}, {\it Phys. Rev. Lett.\/} {\bf 106}, 130506 (2011).
    \bibitem{Friis2018} N. Friis, {\it et~al.\/}, {\it Phys. Rev. X.\/} {\bf 8}, 021012 (2018).  
    \bibitem{DiCarlo2010} L.~DiCarlo, {\it et~al.\/}, {\it Nature\/} {\bf 467}, 574 (2010).  
    \bibitem{Song2017a} C.~Song, {\it et~al.\/}, {\it Phys. Rev. Lett.\/} {\bf 119}, 180511 (2017).  
    \bibitem{Vlastakis2013} B.~Vlastakis, {\it et~al.\/}, {\it Science\/} {\bf 342}, 607 (2013).  
    \bibitem{Labuhn2016} H.~Labuhn, {\it et~al.\/}, {\it Nature\/} {\bf 534}, 667 (2016).  
    \bibitem{Bernien2017} H.~Bernien, {\it et~al.\/}, {\it Nature\/} {\bf 551}, 579 (2017).  
    \bibitem{Jaksch2000} D.~Jaksch, {\it et~al.\/}, {\it Phys. Rev. Lett.\/} {\bf 85}, 2208 (2000).  
    \bibitem{Wilk2010} T.~Wilk, {\it et~al.\/}, {\it Phys. Rev. Lett.\/} {\bf 104}, 010502 (2010).  
    \bibitem{Isenhower2010a} L.~Isenhower, {\it et~al.\/}, {\it Phys. Rev. Lett.\/} {\bf 104}, 010503 (2010).
    \bibitem{Islam2013} R.~Islam, {\it et~al.\/}, {\it Science\/} {\bf 340}, 583 (2013).  
    \bibitem{Rach2015} N.~Rach, M.~M. M\"uller, T.~Calarco, S.~Montangero, {\it Phys. Rev. A\/} {\bf 92}, 062343 (2015).
    \bibitem{Heck2018} R.~Heck, {\it et~al.\/}, {\it PNAS\/} {\bf 115}, E11231 (2018).  
    \bibitem{SOM} See Supplementary Materials
    \bibitem{deLeseleuc2018} S.~{de L\'es\'eleuc}, D.~Barredo, V.~Lienhard, A.~Browaeys, T.~Lahaye, {\it Phys. Rev. A\/} {\bf 97}, 053803 (2018).
    \bibitem{Garttner2017} M.~G\"arttner, {\it et~al.\/}, {\it Nature Physics\/} {\bf 13}, 781 (2017).  
    \bibitem{Farhi2014} E.~Farhi, J.~Goldstone, S.~Gutmann, {\it arXiv:1411.4028\/}  (2014).  
    \bibitem{Zhou2018} L.~Zhou, S.-T. Wang, S.~Choi, H.~Pichler, M.~D. Lukin, {\it arXiv:1812.01041 [cond-mat, physics:quant-ph]\/}  (2018).  
    \bibitem{Turner2018a} C.~J. Turner, A.~A. Michailidis, D.~A. Abanin, M.~Serbyn, Z.~Papi\'c, {\it Nature Physics\/} {\bf 14}, 745 (2018).
    \bibitem{Picken2018} C.~J. Picken, R.~Legaie, K.~McDonnell, J.~D. Pritchard, {\it Quantum Sci.  Technol.\/} {\bf 4}, 015011 (2018).
    \bibitem{Song2019} C.~Song, {\it et~al.\/}, {\it arXiv:1905.00320\/}  (2019).
    \bibitem{Wei2019} K.~X. Wei, {\it et~al.\/}, {\it arXiv:1905.05720\/}  (2019).
\end{thebibliography}

\begin{thebibliography}{20}
    \makeatletter
    \addtocounter{NAT@ctr}{35}
    \makeatother
    \bibitem{Levine2018} H.~Levine, {\it et~al.\/}, {\it Phys. Rev. Lett.\/} {\bf 121}, 123603 (2018).
    \bibitem{Montangero2007} S.~Montangero, T.~Calarco, R.~Fazio, {\it Phys. Rev. Lett.\/} {\bf 99}, 170501 (2007).
    \bibitem{Brif2010} C.~Brif, R.~Chakrabarti, H.~Rabitz, {\it New J. Phys.\/} {\bf 12}, 075008 (2010).
    \bibitem{Cui2015} J.~Cui, F.~Mintert, {\it New J. Phys.\/} {\bf 17}, 093014 (2015).
    \bibitem{Koch2016} C.~P. Koch, {\it J. Phys.: Condens. Matter\/} {\bf 28}, 213001 (2016).
    \bibitem{Doria2011} P.~Doria, T.~Calarco, S.~Montangero, {\it Phys. Rev. Lett.\/} {\bf 106}, 190501 (2011).
    \bibitem{Glaser2015} S.~J. Glaser, {\it et~al.\/}, {\it Eur. Phys. J. D\/} {\bf 69}, 279 (2015).
    \bibitem{Brouzos2015} I.~Brouzos, {\it et~al.\/}, {\it Phys. Rev. A\/} {\bf 92}, 062110 (2015).
    \bibitem{Lloyd2014} S.~Lloyd, S.~Montangero, {\it Phys. Rev. Lett.\/} {\bf 113}, 010502 (2014).
    \bibitem{vanFrank2016} S.~{van Frank}, {\it et~al.\/}, {\it Scientific Reports\/} {\bf 6}, 34187 (2016).
    \bibitem{Muller2013} M.~M. M\"uller, {\it et~al.\/}, {\it Phys. Rev. A\/} {\bf 87}, 053412 (2013).
    \bibitem{Caneva2009} T.~Caneva, {\it et~al.\/}, {\it Phys. Rev. Lett.\/} {\bf 103}, 240501 (2009).
    \bibitem{Caneva2011a} T.~Caneva, T.~Calarco, R.~Fazio, G.~E. Santoro, S.~Montangero, {\it Phys. Rev.  A\/} {\bf 84}, 012312 (2011).
    \bibitem{Cui2017} J.~Cui, R.~van Bijnen, T.~Pohl, S.~Montangero, T.~Calarco, {\it Quantum Sci.  Technol.\/} {\bf 2}, 035006 (2017).
    \bibitem{Nelder1965} J.~A. Nelder, R.~Mead, {\it Comput J\/} {\bf 7}, 308 (1965).
    \bibitem{Kaufman2012a} A.~M. Kaufman, B.~J. Lester, C.~A. Regal, {\it Phys. Rev. X\/} {\bf 2}, 041014 (2012).
    \bibitem{Thompson2013a} J.~D. Thompson, T.~G. Tiecke, A.~S. Zibrov, V.~Vuletic, M.~D. Lukin, {\it Phys.  Rev. Lett.\/} {\bf 110}, 133001 (2013).
    \bibitem{Khemani2019} V.~Khemani, C.~R. Laumann, A.~Chandran, {\it Phys. Rev. B\/} {\bf 99}, 161101 (2019).
    \bibitem{Ho2019a} W.~W. Ho, S.~Choi, H.~Pichler, M.~D. Lukin, {\it Phys. Rev. Lett.\/} {\bf 122}, 040603 (2019).
    \bibitem{Beterov2009a} I.~I. Beterov, I.~I. Ryabtsev, D.~B. Tretyakov, V.~M. Entin, {\it Phys. Rev.  A\/} {\bf 79}, 052504 (2009).
\end{thebibliography}
\end{document}